# Study of Entropy-Diffusion Relation in a Deterministic Hamiltonian System through Microscopic Analysis

**Subhajit Acharya** and **Biman Bagchi***

Solid State and Structural Chemistry Unit

Indian Institute of Science, Bengaluru, India

*corresponding author's email: bbagchi@iisc.ac.in

**Abstract**

*Although an intimate relation between entropy and diffusion has been advocated for many years, and even seems to have been verified in theory and experiments, a quantitatively reliable study, and any derivation of an algebraic relation between the two does not seem to exist. Here we explore the nature of this entropy-diffusion relation in three deterministic systems where an accurate estimate of the both can be carried out. We study three deterministic model systems, (a) the motion of a single point particle with constant energy in a two-dimensional periodic potential energy landscape, (b) the same in regular Lorentz gas where a point particle with constant energy moves between collisions with hard disc scatterers and (c) motion of a point particle among the boxes with small apertures. These models, introduced by Zwanzig, exhibit diffusive motion in the limit where ergodicity is shown to exist. We then explore the diffusion-entropy relation by an accurate calculation of both diffusion and entropy for the aforementioned model systems. We estimate the self-diffusion coefficient of the particle by employing computer simulations and entropy by quadrature using Boltzmann's formula. We observe an interesting crossover in the diffusion-entropy relation in some specific regions which is attributed to the emergence of correlated returns. The crossover could herald a breakdown of the Rosenfeld-like exponential scaling between the two, as observed at low temperatures. Later, we modify the scaling relation to account for the correlated motions and present a detailed analysis of the dynamical entropy obtained via Lyapunov exponent which is rather an important quantity in the study of deterministic systems.*

## I. Introduction

The relationship between entropy (S) and diffusion coefficient (D) is highly intriguing in nature since it relates completely two seemingly unrelated properties namely, structure and dynamics. Diffusion coefficient again is related to the coefficient of friction ($\zeta$) through the well-known Einstein's relation[1] $D = \frac{k_B T}{\zeta}$. This exact relation is often found to be useful widely in the liquid state dynamics,[2] where the friction is found to be proportional to the viscosity of the medium. Alternatively, there is an intimate connection between noise of any system and friction via the fluctuation-dissipation theorem.

On the other hand, a related, highly discussed and debated issue is the oft-quoted relationship between diffusion and entropy. While entropy is a measure of accessible phase space, diffusion gives the rate of exploration of the configuration space. Hence, a scaling relation connecting these two is highly intriguing at a fundamental level. We often find two such well-known universal relations in the literature. First one is the well-known Adam-Gibbs relation[3,4] given by [**Eq.1**]

$$D = A e^{-\left(B/TS_c\right)} \tag{1}$$

where $S_c$ represents configurational entropy per particle of the system at a temperature T; A and B are two arbitrary scaling constants. Configurational entropy can be determined by subtracting vibrational entropy per particle from the total entropy per particle of the system i.e. $S_c(T) = S(T) - S_{vib}(T)$. While this famous relation is found to hold well at a low temperature near the glass transition, its validity at higher temperatures has been questioned.[5–7] The other relationship between entropy and diffusion was proposed by Rosenfeld[8] based on the previous simulation works for the transport coefficients of a wide variety of systems and is given by [**Eq.2**].

$$D^* = a\exp(bS_{ex}) \qquad (2)$$

Here, excess entropy is given by $S_{ex} = \dfrac{S - S_{id}}{Nk_B}$ where, $S$ and $S_{id}$ represent total entropy of the system and ideal gas respectively. $D^*$ denotes the reduced diffusion coefficient and $N$ represents total number of atoms/molecules present in the system. In **Eq.2** $a$ and $b$ are two constants that depend on the nature of the system. While Adam–Gibbs relation was first derived in order to explain the relaxation phenomenon in glassy systems[4,9] introducing the concept of '*cooperatively rearranging region*', Rosenfeld scaling[10–12] is mainly valid at temperatures, higher than the melting temperature where the system is expected to be fully ergodic. But still, there is no precise explanation of the origin of relation between diffusion and entropy. Moreover, the difficulty of establishing a precise D-S relation can be attributed partly to the difficulty in obtaining an *estimation of the entropy explored by the diffusing particle.* This difficulty may be addressed by studying simple deterministic Hamiltonian systems where entropy can be calculated accurately.

In fact, there has also been keen interest on diffusion in simple deterministic Hamiltonian systems[13–15] that can exhibit random motions and can become chaotic even in the absence of noise. Several works[16–18] have been devoted in search for the origin of the transport coefficients like diffusion in these deterministic systems We refer to the seminal works by Buminovich-Sinai[13] and Machta and Zwanzig.[14] Later, Bagchi *et al* [15] studied a periodic analog of regular Lorentz gas but with continuous potential where the potential energy surface in two dimensions is given by [**Eq.3**]

$$V(x,y) = \cos\left(x + \frac{y}{\sqrt{3}}\right) + \cos\left(x - \frac{y}{\sqrt{3}}\right) + \cos\left(\frac{2y}{\sqrt{3}}\right) \tag{3}$$

In this case, dynamics of the point particle are complex due to refocusing caused by a concave curvature in the potential energy surface. Our interest in these systems stems from the fact that one can calculate entropy accurately almost analytically, although its dynamics could rather be complex.

In addition, there are several interesting questions yet to be answered. First, what are the conditions for which the motion of a deterministic system is diffusive (or, ergodic)? It is of course unclear whether diffusion at all could exist for deterministic Hamiltonian systems, without any noise term. Second, is it possible to develop a generalized microscopic theory for establishing exponential scaling relation starting from basic principles of Statistical mechanics? Third, can we find simple deterministic systems where entropy can be calculated accurately? Can we establish any relation, if any, between D and S?

In this work, we inquire the abovementioned interesting questions. The rest of the article is organized as follows. In section II, we present a detailed description of the microscopic derivation of the diffusion-entropy relation. In section III, we describe the deterministic Hamiltonian model systems along with simulation details. Simulation results are presented in section IV. Finally, in section V, we summarize our work and draw some general conclusions.

## II. Theoretical Considerations

Despite great interest[10,11] on the origin of diffusion-entropy relation, a precise derivation of the relation is yet to be established using the principles of Statistical Mechanics. In the earlier derivation of scaling relation by Banerjee *et al*,[11] the double integral expression of mean first time was employed, with certain approximations. We note that there must be an underlying random walk behind the existence of diffusion in any system. In some cases, random walk nature is easy to decipher, like for Lorentz gas at small value of separation parameter[14], but can be hard for some others. The regular random walk model allows us to use the relation between the rate constant ($k$) of crossing from one cell to the other, to the diffusion coefficient(D) through the relation[19] like, $D = \frac{1}{2d}k(E)a^2$ where, $a$ is lattice constant, the distance between two adjacent cells, and $d$ represents the dimension of the system. Again, according to the celebrated transition state theory of Wigner and Eyring[20], the rate constant is a property of phase space trajectories. We use this connection to relate diffusion with entropy.

In the transition state theory formalism[21], the general expression of the energy ($E$) dependent rate constant $k(E)$ in a microcanonical ensemble is given by [22–24]

$$k(E) = \frac{\frac{1}{2}\int \frac{dq^N dp^N}{(2\pi\hbar)^N} \delta(H_N - E)\delta(q_1 - q_c)|\dot{q}_1|}{\int \frac{dq^N dp^N}{(2\pi\hbar)^N} \delta(H_N - E)} \qquad (4)$$

Where, $H_N$ is the total Hamiltonian of the system with $N$ degrees of freedom corresponding to the total energy $E$ and is defined as $H_N = \sum_{i=1}^{N} \frac{p_i^2}{2m} + V(q_1,....q_N)$. $p_i$ denotes the conjugate momentum along the $i^{th}$ degrees of freedom of the particle of mass $m$ and $V(q_1,....q_N)$ is the total potential energy of the system. In the preceding equation, $\dot{q}_1$ defines velocity along the reaction coordinate only.

In the derivation of **Eq.4** we assume that one of the coordinates i.e., reaction coordinate (say, $q_1$) is perpendicular to the dividing surface $\sigma(q_1 = q_c) = 0$ which separates reactant from product side where, $q_c$ represents critical value of $q_1$ at the transition point. Our goal is to get the simplified form of numerator and denominator of **Eq.4**. The denominator of the **Eq.4** represents the classical density of states per unit energy and it is denoted by

$$\rho_c(E) = \frac{1}{h^N} \int dq^N dp^N \delta(H_N - E)$$ .We simplify $\rho_c(E)$ further as follows,

$$\rho_c(E) = \frac{d}{dE} \int \frac{dq^N dp^N}{(2\pi\hbar)^N} \Theta(H_N - E) = \frac{d\Omega(E)}{dE}.$$

(5)

Here, $\Omega(E)$ represents the total number of states with energy less than or equal to $E$ and $\Theta$ is step a function. In order to simplify the numerator of **Eq.4** we assume that reaction coordinate ($q_1$) at a critical point becomes separable from the other degrees of freedom. Therefore, one can expand the potential energy function around $q_1$ at the transition state as

$$V_N(q_1, q_2, \ldots q_N) = E_0 + V_{N-1}(q_2, \ldots q_N)$$

(6)

In **Eq.6**, $E_0$ denotes the bond dissociation energy along $q_1$ and $V_{N-1}$ denotes the potential energy for the remaining *N*-1 degrees of freedom. Use of the **Eq.6** allows us to separate out the integration over $q_1$ and $p_1$ appeared in the numerator of **Eq.4** as follows

$$\frac{1}{2}\left[\frac{1}{2\pi\hbar} \int dE' \int dq_1 dp_1 \delta\left(\frac{p_1^2}{2m} - E'\right) \delta(q_1 - q_c)\left|\frac{p_1}{m}\right| \times \frac{1}{(2\pi\hbar)^{N-1}} \int dq_2 dp_2 \ldots \delta\left(\sum_{i>1}\frac{p_i^2}{2m} + E_0 + V_{N-1} - E + E'\right)\right]$$

We then perform the integration over $q_1$ and $p_1$ by using two delta functions involved in the preceding equation to obtain

$$\frac{1}{2\pi\hbar} \times \frac{1}{(2\pi\hbar)^{N-1}} \int dq^{N-1} dp^{N-1} \int dE' \frac{d}{dE'} \Theta(H_{N-1} - (E - E_0) + E')$$

$$= \frac{1}{2\pi\hbar} \Omega^{tran}$$

(7)

Where, $\Omega^{tran}(E) = \frac{1}{(2\pi\hbar)^{N-1}} \int dq_2 dp_2 \ldots \int dE' \frac{d}{dE'} \Theta\left(\sum_{i>1}\frac{p_i^2}{2m} + E_0 + V_{N-1} - E + E'\right).$

We invoke the simplified form of numerator and denominator i.e., **Eq.7** and **Eq.5** in **Eq.4** to obtain, $k(E) = \dfrac{k_B}{2\pi\hbar \left(\dfrac{\partial S}{\partial E}\right)} \dfrac{\Omega^{tran}(E)}{\Omega(E)}$. This form of the equation is analogous to Wigner[20] and Eyring[25,26] equation obtained from transition state theory approximation if we use the definition of temperature in microcanonical ensemble i.e. $\left(\dfrac{\partial S}{\partial E}\right)_{N,V} = \dfrac{1}{T}$. Using the relation between D and $k(E)$ mentioned above and introducing $S^{\dagger} = k_B \ln \Omega^{tran}$ (i.e. Boltzmann's formula of entropy[27]) we recover the following interesting expression for the diffusion constant

$$D = \dfrac{a^2}{2d} \dfrac{k_B}{2\pi\hbar \left(\dfrac{\partial S}{\partial E}\right)} \exp\left(-\dfrac{S - S^{\dagger}}{k_B}\right) \quad (8)$$

**Eq.8** is new. It is similar to the form of Rosenfeld, but not identical. *Here $S^{\dagger}$ is a scaling constant that serves the same role of $S_{id}$ involved in Rosenfeld scaling relation (Eq.2). In the latter, not only the ideal gas entropy term, but also the entire form was adopted phenomenologically.*

### III. Systems and Simulation details

We study three deterministic systems where entropy and diffusion can be calculated accurately namely,(a) the motion of a single point particle with constant energy in a two-dimensional periodic potential energy landscape defined by sum of cosine potentials as shown in **figure 1b**, (b) the same in a regular Lorentz gas [**figure 1a**] and (c) the same among the boxes with small apertures as shown in **figure 1c**. Study of diffusion in periodic cosine potential and Lorentz gas is otherwise also important due to the existence of a variety of applications like super-ionic conductors, motion of adsorbates on crystal surfaces, polymers diffusing at the interfaces, molecular graphene, etc.[28–33]

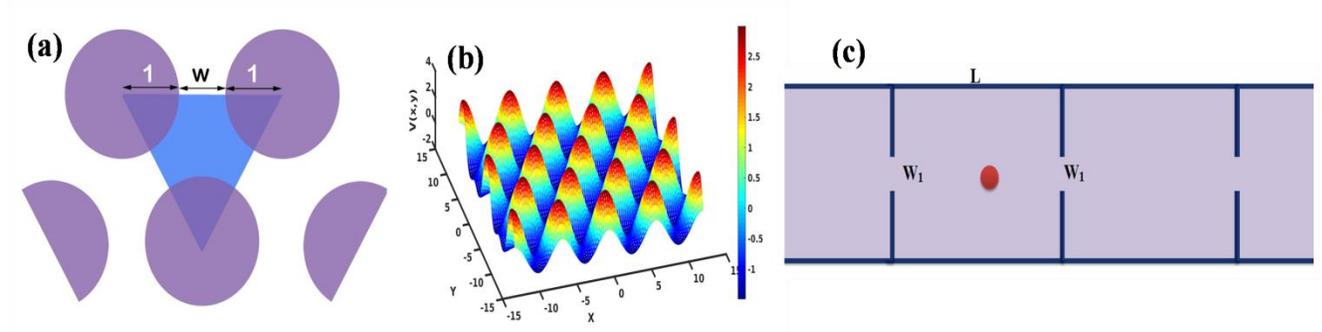

Figure 1:(a)The location of fixed hard-scatterers in a periodic Lorentz model. The blue shaded region denotes a trapping region of triangular symmetry. W is the separation parameter between hard disks. (b) A schematic representation of potential energy function defined in Eq.3. The range and color codes are given on the right side of the plot. (c) A schematic diagram of box-hole model where the point particle (i.e., grey-colored circle as shown in the figure) moves among the boxes with small holes of width $W_1$ allowing long distance motion only in one direction. In (c) L indicates the length of the box.

We next project potential function defined by **Eq.3** into the x-y plane to find that each cell contains a minimum at the center of the triangle, maxima at the three corners of the triangle and saddle points at the mid-points of the edges. Energies of the maxima, minimum and saddle points of each cell are given by 3.0,-1.5 and -1.0 respectively. The point particle can explore the whole phase space if its energy is higher than the saddle point energy. Otherwise, it can get trapped inside a cell forever. We start our simulation in constant NE ensemble placing the particle near the minima of a cell. We use Gear's fifth order predictor-corrector algorithm[34] for integrating the equation of motion of the particle with timestep 0.001. We perform simulations for a set of different values of the total energy of the system for $10^6$ steps. Each of the calculations is performed for 100 different initial configurations in order to obtain a statistically significant and reproducible result.

In the case of Machta – Zwanzig (MZ) model we exactly follow the outlines mentioned in the classic article by Zwanzig *et al*[14] to simulate a point particle moving in between the triangular array of static hard disc scatterers. We assume the radii of the scatterer to be one and spacing between them is to be *W* as shown in **Figure 1a**. We perform simulation of a point particle starting from different positions for the several values of separation parameter.

We study the box-hole model system also apart from continuous periodic potential and periodic Lorentz model. **Figure 1c** shows the schematic representation of this model. Initially, the point particle starts from an arbitrary position of any box with side length 'L'. While exploring the phase space it suffers an elastic collision with the wall of the box. The Motion of the point particle is bounded by the wall of the box in the vertical direction only. But, the particle can move to its adjacent box through a hole of width $W_1$ present in the horizontal directions as shown in **Figure 1c**. Like the Lorentz model here also the static and dynamic properties are calculated for a set of values of $W_1$.

## IV. Results & Discussions

### A. Diffusion-entropy Relation for the Deterministic systems

In this section we explore the exponential relations between diffusion and entropy for the following deterministic systems.

#### a. For continuous periodic potential :

In this case, dynamics of the point particle are complex due to refocusing caused by a concave curvature in the potential energy surface (as shown in **Figure 1b**).This causes the trajectories of the particle to be trapped during different back-and-forth journeys between the same two cells. The existence of diffusion is the consequence of these multiple collisions with the potential energy surface. The self-diffusion coefficient (D) of the point particle is calculated using mean square displacement (MSD) and velocity auto-correlation function (VACF). In two-dimension self-diffusion coefficient is defined by

$$D = \lim_{t \to \infty} \frac{\left\langle (r(t) - r(0))^2 \right\rangle}{4t} \qquad (9)$$

Where, $r(t)$ is the position of the particle at time $t$ and angular brackets indicate the ensemble average. Here, we compute the MSD $\left\langle (r(t) - r(0))^2 \right\rangle$ by taking an average of over 100 trajectories with completely different initial configurations. Alternatively, the self-diffusion coefficient can be calculated by integrating the un-normalized velocity autocorrelation. According to the Green-Kubo formalism, D is defined as,

$$D = \frac{1}{d} \int_0^\infty \langle v(t) \cdot v(0) \rangle dt \qquad (10)$$

Where, $d$ indicates the dimension of the system and $v(t)$ is the velocity vector of the particle at time $t$. The variations of MSD and un-normalized VACF with time for different values of energy are shown in **Figure 2(a)** and **Figure 2(b)**.

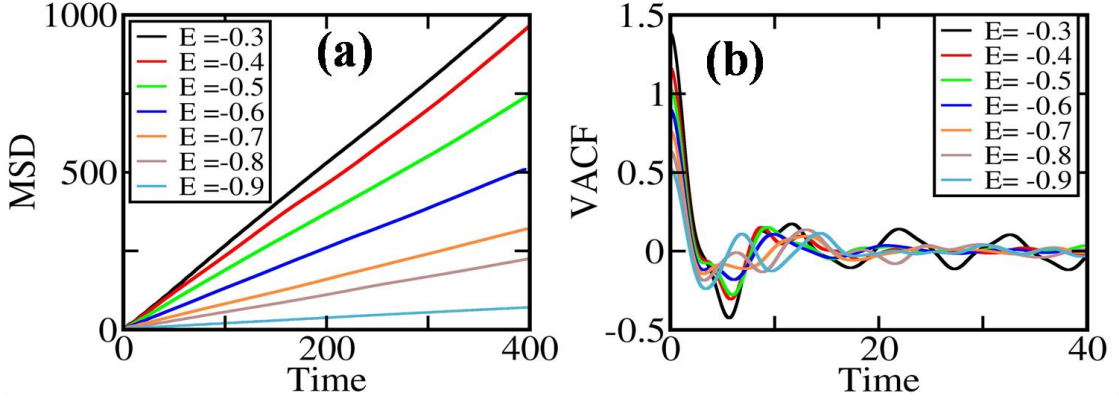

**Figure 2:** (a) variations of MSD against time (b) variations of un-normalized VACF against time for different values of energy of the system. We take an ensembleaverage over 100 different trajectories for both MSD and VACF.

Both methods give similar values of D. VACF exhibits negative region because of trapping but decays at a faster rate with increasing energy up to a certain critical energy. We discover the critical value of energy (to be calculated -0.4) beyond which the relaxation of VACF becomes more oscillatory in the long-time limit. Hence, above the critical energy, VACF could not provide a well-defined self-diffusion coefficient for this system. In **Figure 3a**, we plot D obtained from MSD against energy.

The next turn is the calculation of entropy accurately. We use Boltzmann's formula, $S = k_B \ln \Omega$ to calculate the entropy, where $\Omega$ represents partition function in the micro-canonical ensemble. For our system, we solve the following integration numerically in order to get an estimation of $\Omega$ for a particular cell using conventional quadrature methods,**[Eq.11]**

$$\Omega = \frac{1}{c} \int dx \int dy \int dp_x \int dp_y \, \delta(H - E) \quad (11)$$

Here, $E$ represents the total energy, $c$ denotes normalization constant and $H$ corresponds to the total Hamiltonian of the continuous potential (i.e., **Eq.3**) that is given by $H = \frac{p_x^2}{2m} + \frac{p_y^2}{2m} + V(x, y)$. As discussed before, $p_x$ and $p_y$ represent the x and y-components of the momentum of the particle respectively and *m* is the mass of the particle which is taken as unity in this case. Similarly, we compute the entropy of the ideal system by considering constant flat potential energy surface by carrying out the integration as follows, $\Omega^{id} = A/c \int dp_x \int dp_y \delta\left(\frac{p_x^2}{2m} + \frac{p_y^2}{2m} - E'\right)$. *A* is the area of the triangle and $E'$ indicates the constant energy of the system because the collisions are elastic and *c* is normalization constant as usual. We define scaled entropy ($S_1$) of the form $S_1 = k_B \ln \int dx \int dy \int dp_x \int dp_y \delta(H-E)$ and variation of $S_1$ against energy is plotted in **Figure 3b**. In **Figure 3c** we plot D against the excess entropy for different values of energy of the system. From **Figure 3c** it is noted that exponential[10] relation holds remarkably well with two distinct exponents indicating the presence of two distinct regimes. Therefore, **Figure 3(c)** can be treated as the signature of a crossover at E=-0.9 from low energy, quasi-ergodic to ergodic. The correlated random walks which appear due to the back and forth motions between the cells start to dominate near saddle energy(i.e. -1.0) and playing key role behind the breakdown of regular random walk model. We shall discuss the details in the subsequent sections.

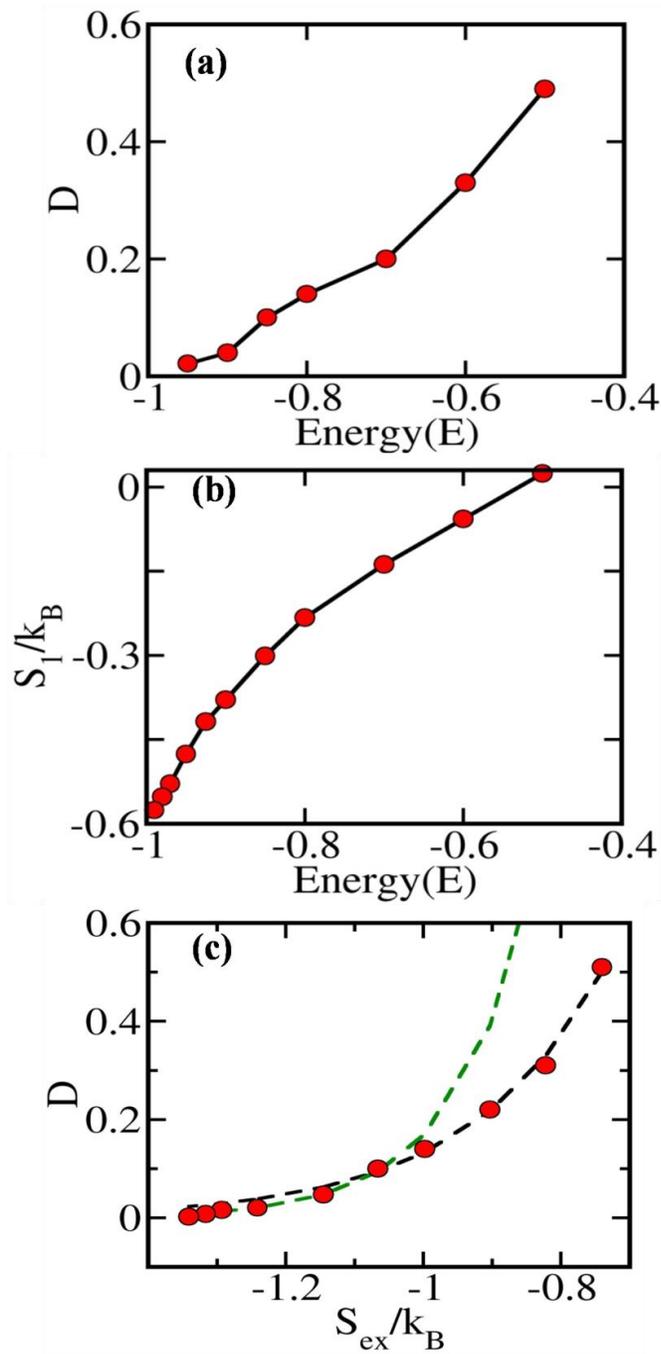

**Figure 3:** **(a) plot of D against energy (E),(b) plot of scaled entropy $S_1$ against energy and (c) plot of D against excess entropy. In (c) we use two exponential functions of the form $D = a\exp(bS_{ex})$ for fitting the entire data points (black trace and green trace). This plot(c) resembles well-known exponential relation (Eq.2) with a=25.91 and b =5.32 for black trace. In (a) and (b) the lines joining the data points are provided as a guide to the eyes**

### b. For regular Lorentz gas

For Machta-Zwanzig model,[14] D can be estimated analytically as a function of the width of exit($W$) assuming that the regular random walk model for two-dimensional lattice is valid for all values of W and is given by,

$$D(W) = \left(W/\pi\right)(2+W)^2 \left[\sqrt{3}(2+W)^2 - 2\pi\right]^{-1} \quad (12)$$

In **Eq.12**, *W*, the separation parameter, alone determines the behavior of the system as the speed is constant. The self-diffusion coefficient values predicted by the **Eq.12** are plotted in **Figure 4a** as the dotted lines. The red circles connected by the line in **Figure 4a** represent our simulation results for self diffusion coefficient obtained from simulation via MSD. **Figure 4a** shows the marked deviation of simulation results from the analytical approximated results in the regions 0<W<0.1 and W> 0.2. The origin of the significant deviations is well-known in literature[35] and is mainly attributed to correlated dynamics, backscattering probability etc. Our goal is to study the dynamics of point particle for values of separation parameter (W) close to zero.
We derive an exact expression of Boltzmann entropy for the system which is given as

$$S(W) = k_B \ln \frac{2\pi}{c} \left[\frac{\sqrt{3}}{4}(2+W)^2 - \frac{\pi}{2}\right] \quad (13)$$

Where, *c* represents normalization constant. While deriving **Eq.13** we assume radii of the scatter to be unity. In order to get an ideal limit of *S(W)* we neglect interaction effect by assuming scatterers as point particles.

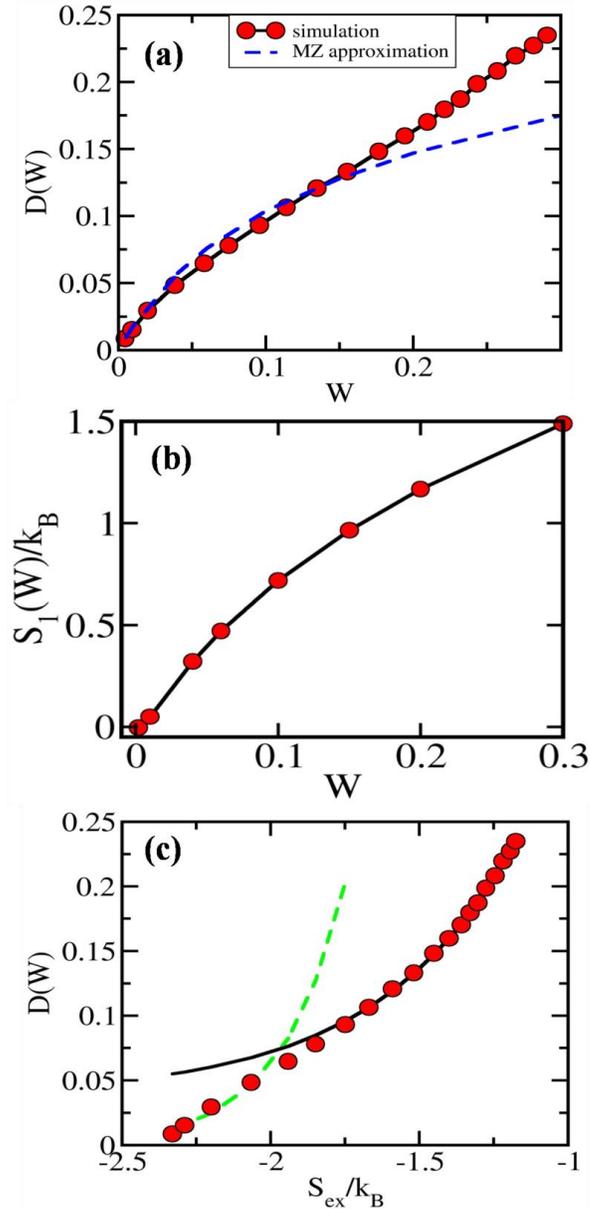

**Figure 4:** (a) plot of D against separation parameter (W). The black line connecting red circles represents the variations of D obtained from computer simulation by means of MSD. In the plot, blue dotted line indicates the variations of D in accordance with Eq.12 which is rather known as Machta-Zwanzig (MZ) approximation (b) plot of scaled Boltzmann entropy against W and (c) plot of D against excess entropy. Here in (c), we use two exponential functions of the form $D = a exp(bS_{ex}) + c_1$ for fitting the data sets(black trace and green trace) over all values of W. This plot (i.e., black trace) agrees with Rosenfeld scaling (Eq.2) with a=2.47 and b=0.46. In (a) and (b) the lines joining the data points are provided as a guide to the eyes.

We plot the scaled entropy against W in **Figure 4b**. In **Figure 4c**, we plot D obtained via computer simulation against excess entropy for a particular triangular trap and we find that two exponential functions are needed (black and green trace) to fit the entire region of separation parameter. We observe an interesting crossover in the diffusion-entropy scaling plot near the region $W \rightarrow 0$ which is also attributed to the emergence of pronounced correlated random returns. We shall present a detailed discussion on the modified rate constant influenced by the pronounced correlated random walks later.

### c. Box-hole model

For the box-hole model we estimate the self-diffusion coefficient obtained from computer simulation by means of MSD for different values of $W_1$. We plot D obtained against $W_1$ in **Figure 5a**.

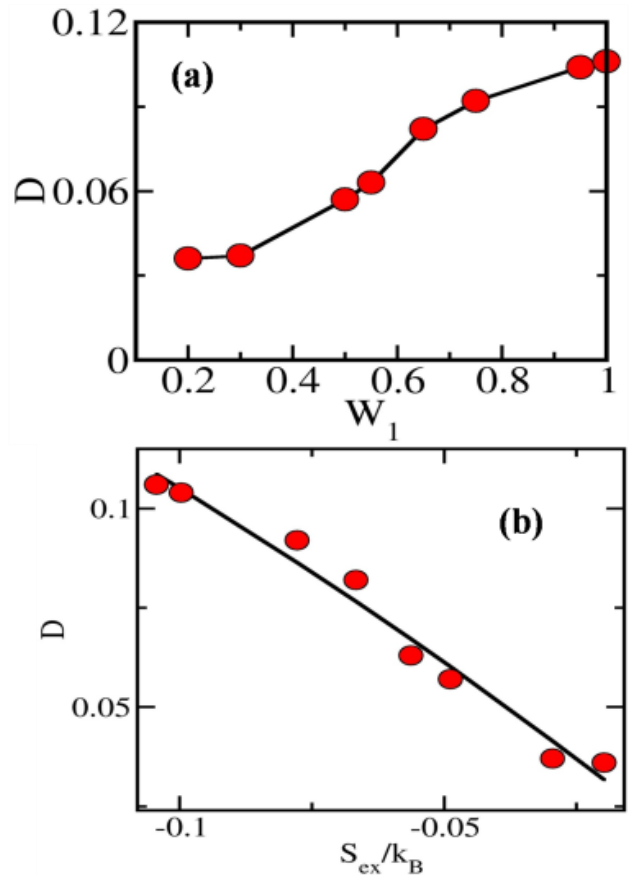

**Figure 5:** (a) plot of self-diffusion coefficient against $W_1$. Here, the line joining the data points is provided as a guide to the eyes. (b) plot of D against excess entropy. In (b) we use $D = a\exp(bS_{ex}) + c_1$ for fitting the data points (black trace). This plot (b) resembles well-known Rosenfeld scaling with a=-0.38 and b =0.36.

Like the periodic Lorentz model we use the following relation in order to calculate excess entropy for this system for different values of $W_1$,

$$S_{ex} = S - S_{id} = -k_B \ln\left(1 + \frac{1}{L/W_1 - 1}\right) \qquad (14).$$

In **Figure 5b** we show the variation of D against excess entropy for different values of $W_1$. From **Figure 5b**, it is noted that exponential relation between D and S holds for this system remarkably.

## B.    Dynamical entropy in case of periodic potential

For a deterministic system, dynamical randomness is generally characterized by a positive value of Kolmogrov-Sinai entropy[36–38] ($h_{KS}$). A measure of the onset of chaos or dynamical randomness[39] can be obtained from the sensitivity to the initial conditions in a closed dynamical system which is again characterized by Lyapunov exponent.[40,41] In a two-dimensional Hamiltonian system, the dynamical entropy can be determined by the following formula $h_{KS} = L_n \mu$, where $L_n$ is the maximal average,[42] LCE as discussed later. $\mu$, an important characteristic of the chaotic motion, is defined as $\mu = \frac{n_{ch}}{n}$ where $n$ indicates the total number of grid points accessible to our point particle and $n_{ch}$ denotes the number of grid points responsible for the chaotic motion for a particular triangular cell. In **Figure 6a** we plot $\mu$ against energy. From the figure we note that $\mu$ increases with the increase in energy as expected. In **Figure 6b** we show the variation of $h_{KS}$ with energy and plot of D against $h_{KS}$ is shown in **Figure 6c**. Interestingly, we find that D exhibits exponential dependence on $h_{KS}$ like Boltzmann entropy since a linear relation holds between Boltzmann entropy and $h_{KS}$ in our system as shown in **Figure 6d**.

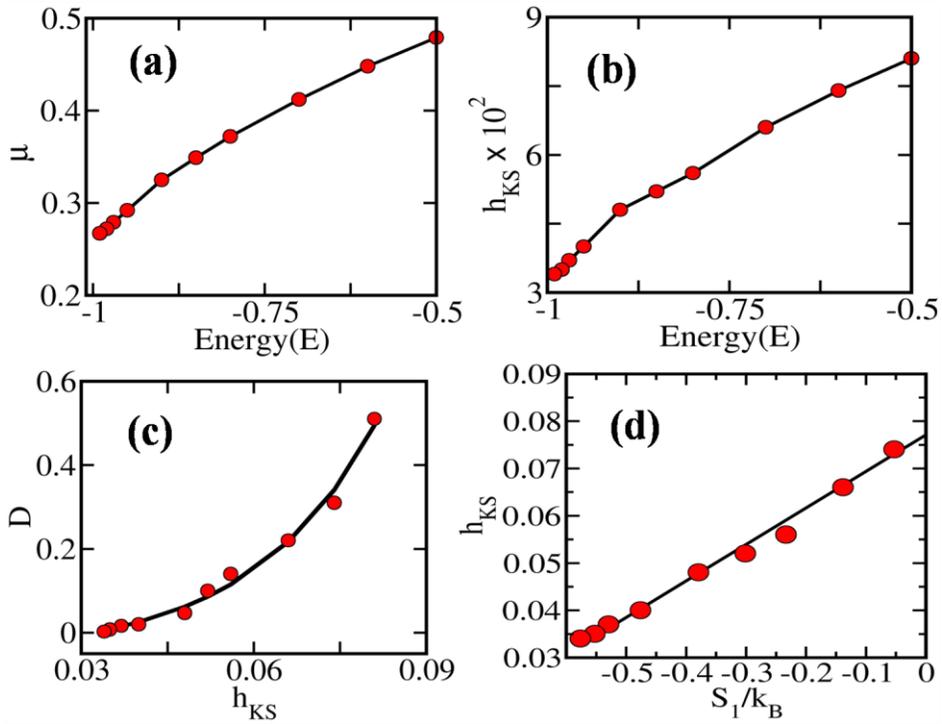

**Figure 6:** For the system with periodic potential (a) Plot of relative area($\mu$) occupied by the chaotic component against E. (b) variation of $h_{KS}$ against energy (c) plot of D obtained from MSD against $h_{KS}$. In (c) we use exponential function $D = a\exp(bh_{KS})$ for fitting (black trace) with a=0.003 and b=62.56. (d) plot of Sinai entropy against $S_1/k_B$ for different values of energy. We use linear function of the form $h_{KS} = a_0 + a_1 S_1/k_B$ for fitting the data points (black trace) with $a_0 = 0.07, a_1 = 0.08$.

## C. Lyapunov exponent for periodic potential and origin of crossover

In order to understand the spread of the trajectories, Lyapunov exponents[43] are calculated Chaotic behavior can be generated within a dynamical system even in the absence of external random forces. Quantitatively this chaotic motion can be determined by computing Lyapunov characteristic exponent (LCE).[44] LCE of a trajectory basically determines the average exponential rate of convergence or divergence of other trajectories surrounding it in phase space. LCE is generally zero for regular motions and takes positive value for chaotic motions in the trajectory. It is found that computing maximum LCE ($L_n$) is sufficient in order to characterize the motion of the system. We use 'shadow trajectory' method[42] in order to compute $L_n$ and it is defined as follows,[Eq.15]

$$L_n = \frac{1}{n\Delta t} \sum_{i=1}^{n} \ln \frac{d_i}{d_{i-1}} \qquad (15)$$

In this method, two arbitrary close points, one corresponding to the reference trajectory and another to a shadow trajectory in the phase space are chosen and initial separation between them ($d_0$) is computed. In the above expression (**Eq.15**) $d_i$ represents a separation between reference and shadow trajectory at $i^{th}$ step and $L_n$ denotes maximal LCE at the $n^{th}$ step with time interval $\Delta t$. During calculation, we periodically renormalize the separation between the reference and the shadow trajectories. In **Figure 7a** dependence of $L_n$ on different values of energy is shown. Clearly, $L_n$ increases with energy as motion of the particle becomes chaotic at a faster rate.

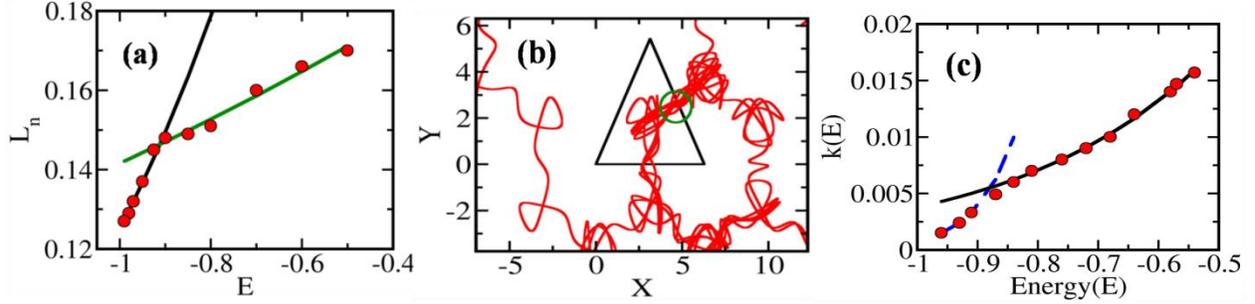

**Figure 7**: (a) Plot of $L_n$ against energy. We observe a clear crossover near E=-0.9 that indicates the presence of distinctly different dynamics in the two regimes. Here, line joining the data points (black and green trace) are provided as a guide to the eyes. (b) Plot of phase space trajectory of the point particle moving with constant energy. Central black triangle represents the initial cell of the particle. We observe multiple crossing (green circled region) through the saddle surface indicating the signature of correlated motions. (c) plot of energy dependent rate constant k(E) against energy (E). Here, we use $D = \frac{1}{4}k(E)a^2$ to obtain $k(E)$ using the values of D obtained from simulation. Clearly, there is a crossover in the plot near the saddle point energy indicating the breakdown of regular random walk model. Figure 7(a), (b) and (c) are for periodic potential (Eq.3).

**Figure 7a** also represents an indication of crossover near the saddle apart from **Figure 3c and 4c**. In case of periodic potential as we go near the saddle by lowering the total energy of the system, the energy dependent width of exit window gradually decreases. In analogy with the periodic potential, for the periodic Lorentz gas with the decrease of size of exit window (W) we observe a significant deviation in the diffusion-entropy scaling plot. In **Figure 7c,** we plot energy dependent rate constant ($k(E)$) against energy($E$) considering that simple random walk model is valid over the entire energy spectrum. The presence of crossover in this plot is clearly an indication of breakdown of regular random walk model near the saddle point energy. The physical mechanism behind the crossover is the onset of correlated motion that makes the random walk trace the same path repeatedly, by recrossing among the triangular cells(**Figure 7b**). This makes the expression of rate constant (**Eq.4**) used to derive the exponential scaling relation invalid.

In the presence of correlated random walk, **Eq.4** can be modified[15,45] as

$$k(E) = \langle J_e(S) \rangle_e + \int_0^{T_M} dt \langle J_e(S) J_R(S,t) \rangle_e \tag{16}$$

where, $J_e(S)$ and $J_R(S,t)$ denote the outgoing flux crossing the saddle surface S at time t=0 and intrinsically negative flux coming back to the initial surface at later *t*. *The second term is negative, thus lowering the value of the rate constant, and hence of the diffusion coefficient.* Here, the subscript 'e' indicates the equilibrium average in the microcanonical ensemble and $T_M$ is the upper limit of integration in order to avoid the contribution coming from the long time 'thermalized' returns. The transition state theory rate constant (**Eq.4**) becomes equal to $\langle J_e(S) \rangle_e$ in the absence of correlated returns. We can further simplify the **Eq.16** as

$$k(E) = \langle J_e(S) \rangle_e \left[ 1 - \int_0^{T_M} P_R(t) dt \right] \tag{17}$$

In **Eq.17**, $P_R(t)$ indicates the conditional probability of return at time *t* to the starting cell of the point particle provided it crossed the saddle boundary at *t=0* to enter the neighbor cells. The probability of correlated returns becomes more pronounced while the energy of the point particle approaches to the saddle point energy (i.e, -1.0) or in the $W \to 0$ limit. *The appearance of crossover in the diffusion-entropy scaling plot with two distinct exponents is the consequence of correlated returns* as shown in **Figure 7b**. *The presence of crossover is clearly an indication of two distinct dynamical regimes present in the system.*[46,47] The appearance of correlated motions purely originates from the characteristics nature of potential energy surface. The identification of complex dynamics through 'caging' (**Figure 7b**) has made a deep connection with slow glassy liquids in the context of inherent structure analyses of liquids.[48,49]

### D. Time correlation function for periodic potential

An important dynamical quantity is the exit time dynamics from (or, the residence time of) the starting cell. We define two correlation functions $C_S(t)$ and $C(t)$ for the original cell which are defined as follows,[**Eq.18**]

$$C_S(t) = \frac{\langle H'(0)H'(t)\rangle}{\langle H'(0)\rangle} \text{ and } C(t) = \frac{\langle h'(0)h'(t)\rangle}{\langle h'(0)\rangle}$$

(18)

In the definition of $C_S(t)$, $H'(t)$ is a Heaviside function such that, $H'(t)$ is unity as long as the particle resides inside the original cell. But after leaving the original cell for the first time $H'(t)$ becomes 0 for all later t. $H'(0)$ is taken as unity since particle always resides inside the starting cell at *t=0*. However, in the definition of *C*(t), $h'(t)$ ,another Heaviside function, is always one as long as the particle is inside the original cell even after coming back to the original cell and becomes zero whenever the particle escapes outside the initial cell. As before, $h'(0)$ is unity in this case also. Variations of $C_S$(t) and *C*(t) with time for different values of energy are shown in **Figure 7(a)** and **7(b)**.

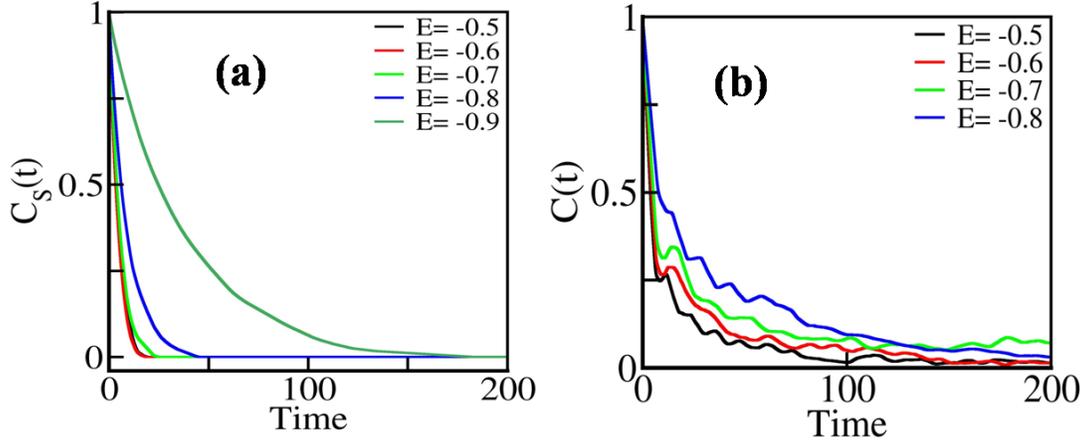

**Figure 8: (a) plot of time-correlation function [$C_S(t)$] for different values of energy. (b) plot of another time correlation function[C(t)] defined by Eq.18 for different values of energy. Plot corresponding to E=-0.9, is highly oscillating due to trapping near minima, is not shown in (b). We note that C(t) decays slowly compared to $C_S(t)$ for a particular value of energy.**

These two functions are mainly introduced in order to capture the short to intermediate dynamics associated with the motion of the point particle moving in continuous potential. We find that for the particular value of energy, $C_S(t)$ exhibits faster decay compared to $C(t)$ since the particle requires a longer time to escape from the original cell permanently. We also note that the plots corresponding to higher energy decay at a faster rate for both the cases since escaping probability from the original cell become more pronounced with increasing energy.

The time correlation function $C_S(t)$ gives an estimate of the exit time of the point particle from the initial cell for the first time. The time decay of correlation function $C_S(t)$ involves two important time scales namely, (i) the time taken to reach the saddle point boundary starting in the vicinity of the minima of the original cell, and (ii) the time required to return to the original cell through the saddle point from the neighboring cell. Hence, we fit the time correlation function $C_S(t)$ using the bi-exponential function of the form $a_1 \exp\left(-\frac{t}{\tau_1}\right) + a_2 \exp\left(-\frac{t}{\tau_2}\right)$. We report fitting parameters in **Table 1.**

**Table 1 : We employ the bi-exponential functions of the form $a_1 exp\left(-\frac{t}{\tau_1}\right) + a_2 exp\left(-\frac{t}{\tau_2}\right)$ to fit the aforementioned time correlation function [$C_S(t)$] against time as shown in Figure 8a. We use the expression $(a_1\tau_1 + a_2\tau_2)$ in order to obtain $\langle\tau\rangle$ i.e. average time required by the particle to escape the original cell. Clearly, crossing the barrier becomes more feasible with the increase in energy of the particle.**

| Energy | $a_1$ | $\tau_1$ | $a_2$ | $\tau_2$ | $\langle\tau\rangle$ |
|---|---|---|---|---|---|
| -0.5 | 0.52 | 4.46 | 0.48 | 4.55 | 4.50 |
| -0.6 | 0.52 | 5.01 | 0.48 | 4.85 | 4.93 |
| -0.7 | 0.40 | 5.46 | 0.60 | 5.43 | 5.44 |
| -0.8 | 0.05 | 3.83 | 0.95 | 9.35 | 9.07 |
| -0.9 | 0.70 | 38.46 | 0.30 | 33.33 | 36.92 |

From the figure we note that crossing the barrier near saddle point becomes more feasible with the increase in energy of the particle. When the energy of the particle is close to saddle point energy (i.e, -1.0) particle takes a much longer time to escape outside the original cell for the first time because of trapping near the minima of the cell.

## V.    Summary & conclusions

Indeed, the existence of diffusion in a deterministic system by itself is a non-trivial issue and a subject of great interest in the academic community. According to hydrodynamic mode-coupling theory,[50] diffusion does not exist in two-dimensions because of its logarithmic diverging nature in the long time limit. But, the origin of the existence of diffusion in our system is something different and discussed below. The most important result is that all the studied systems obey diffusion-entropy relation. Several concluding comments are in order.

(i) We present a general derivation of diffusion-entropy scaling relation for the first time. In the present derivation, we avoid all pitfalls and it is exact. We believe the present derivation will go a long way to explain the 'universality' of the scaling relation.

(ii) Diffusion can never exist in any system without the presence of chaos. On the other hand, chaos is sensitive not only to the initial conditions but also to the characteristics of the potential energy landscape. Several recent studies have revealed a curvature dependent diffusion coefficient for regular Lorentz model motivated by the study of MZ. In case of the periodic Lorentz gas because of the defocusing character of collisions of the point particle with the hard disks the system exhibits Lyapunov instability. The energy landscape must be dispersive. This is satisfied for the periodic Lorentz gas, at the high density of the scatterers. In this case the convexity of the surface of disk scatterers disperses two neighboring trajectories in the phase space.

(iii) However, the motion of the point particle on the triangular cosine potential energy surface is more complex. There are three concave regions in each site that focus trajectories back to the configuration line between the two minima of the adjoining cells. However, existence of sufficient configuration space with dispersive character leads to the existence of diffusive motion.

(iv) The long time rate of dispersion of initially close trajectories in the phase space can be quantified by calculating LCE. The latter provides the rate of exponential growth of separation in the phase space of two initially close trajectories. LCE is supposed to provide a measure of the time of the system to be chaotic. We find that system takes less time to be chaotic with the increase in energy, which is of course expected.

(v) We find an energy dependent crossover in the motion of the particle at low energy when the particle exhibit repeated correlated crossings and re-crossings in the energy landscape. This gives rise to a deviation from exponential diffusion-entropy scaling, and seem to provide an explanation of the breakdown of this scaling at low energies.

(vi) We also estimate the degree of randomness of a time process by evaluating entropy per unit time which is well-known as Kolmogrov-Sinai entropy, or dynamical entropy, of the system for different values of energy.

(vii) We introduce two timescales namely, permanent and mean first exit time of the particle from the starting cell from $C(t)$ and $C_S(t)$ functions that capture the short-to-intermediate dynamical features of the motion of the particle.

## Acknowledgement

We thank Mr. Saumyak Mukherjee and Mr. Sayantan Mondal for discussions. BB thanks Sir J.C. Bose Fellowship and Department of Science and Technology, India (DST) for partial support and SA thanks IISc for the scholarship.

## References

[1] A. Einstein, Ann. Phys. **17**, 549 (1905).

[2] B. Bagchi, *Molecular Relaxation in Liquids* (Oxford Publications, 2012).

[3] G. Adam and J.H. Gibbs, J. Chem.Phys. **43**, 139 (1965).

[4] X. Xia and P.G. Wolynes, Phys. Rev. Lett. **86**, 5526 (2001).

[5] S. Sastry, Nature **409**, 164 (2001).

[6] A. Banerjee, S. Sengupta, S. Sastry, and S.M. Bhattacharyya, Phys. Rev. Lett. **113**, 225701 (2014).

[7] T.R. Kirkpatrick, D. Thirumalai, and P.G. Wolynes, Phys. Rev. A **40**, 1045 (1989).

[8] Y. Rosenfeld, Phys. Rev. A **15**, 2545 (1977).

[9] N. Giovambattista, S. V Buldyrev, F.W. Starr, and H.E. Stanley, Phys. Rev. Lett. **90**, 085506 (2003).

[10] K. Seki and B. Bagchi, J. Chem. Phys. **143**, 194110 (2015).

[11] S. Banerjee, R. Biswas, K. Seki, and B. Bagchi, J. Chem. Phys. **141**, 124105 (2016).

[12] A. Banerjee, M.K. Nandi, and S.M. Bhattacharyya, J .Chem .Sci. **129**, 793 (2017).

[13] L.A. Bunimovich and Y.G. Sinai, Common. Math. Phys. **78**, 247 (1980).

[14] J. Machta and R. Zwanzig, Phys. Rev. Lett. **50**, 1959 (1983).

[15] B. Bagchi, R. Zwanzig, and M.C. Marchetti, Phys. Rev. A. **31**, 892 (1985).

[16] G. Knight, O. Georgiou, C. Dettmann, and R. Klages, Chaos **22**, 023132 (2012).

[17] R. Klages, S. Selene, G. Gallegos, J. Solanpää, M. Sarvilahti, and E. Räsänen, Phys. Rev. Lett. **122**, 64102 (2019).

[18] S. Gil-Gallegos, R. Klages, J. Solanpaa, and E. Rasanen, Eur.Phys.J.Special Top. **228**, 143 (2019).

[19] S. Chandrasekhar, *Rev.mod.Phys.***15**(1),1 (1943).

[20] E. Wigner, Trans. Faraday Soc. **34**, 29 (1938).

[21] P. Pechukas, Ann. Rev. Phys. Chem. **32**, 159 (1981).

[22] H.W. Schranz, L.M. Raff, and D.L. Thompson, Chem.Phys.Lett. **171**, 68 (1990).

[23] A. Lohle and J. Kastner, J. Chem. Theory Comput **14**, 5489 (2018).

[24] J.D. Doll, J. Chem.Phys. **73**, 2760 (1980).

[25] H. Eyring, J. Chem.Phys. **3**, 107 (1935).

[26] K.J. Laidler, *Chemical Kinetics* (Pearson Education, 1987).

[27] L. Boltzmann, *Lectures on Gas Theory* (Dover Publications, 1964).

[28] V. V. Ignatyuk, J. Chem.Phys. **136**, 184104 (2012).

[29] K.J. Challis and M.W. Jack, Phys. Rev. E. **87**, 052102 (2013).

[30] E. Hershkovitz, P. Talkner, E. Pollak, and Y. Georgievskii, Surf. Sci. **421**, 73 (1999).

[31] P.R. Kole, H. Hedgeland, A.P. Jardine, W. Allison, J. Ellis, and G. Alexandrowicz, J. Phys.Condens .Matter. **24**, 104016 (2012).

[32] E. Rasanen, C.A. Rozzi, S. Pittalis, and G. Vignale, Phys. Rev. Lett. **108**, 246803 (2012).

[33] M. Gibertini, A. Singha, V. Pellegrini, M. Polini, G. Vignale, A. Pinczuk, L.N. Pfeiffer, and K.W. West, Phys. Rev. E. **79**, 241406 (2009).

[34] D. Beeman, J. Comput. Phys. **20**, 130 (1976).

[35] R. Klages and C. Dellago, J. Stat.Phy **101**, 145 (2000).

[36] R. Kosloff and S.A. Rice, J. Chem.Phys. **74**, 1340 (1981).

[37] M. Dzugutov, E. Aurell, and A. Vulpiani, Phys. Rev. Lett. **81**, 1762 (1998).

[38] G. Benettin, L. Galgani, and J. Strelcyn, Phys. Rev. A. **14**, 2338 (1976).

[39] J.R. Green, A.B. Costa, B.A. Grzybowski, and I. Szleifer, Proc. Natl. Acad. Sci. U.S .A. **110**, 16339 (2013).

[40] H.V. Beijeren and J.R. Dorfman, Phys. Rev. Lett. **74**, 4412 (1995).

[41] P. Gaspard and G. Nicolis, Phys. Rev. E. **65**, 1693 (1990).

[42] I.I. Shevchenko and A. V Melnikov, JETP Lett. **77**, 642 (2003).

[43] A. Wolf, J.B. Swift, H.L. Swinney, and J.A. Vastano, Physica. **16D**, 285 (1985).

[44] T. Hofmann and J. Merker, CMST. **24**, 97 (2018).

[45] S.H. Northrup and J.T. Hynes, J. Chem.Phys. **73**, 2700 (1980).

[46] P.G. Debenedetti and F.H. Stillinger, Nature **410**, 259 (2001).

[47] S. Sastry, P.G. Debenedetti, and F.H. Stillinger, Nature **393**, 554 (1998).

[48] S. Sastry, P.G. Debenedetti, F.H. Stillinger, J.C. Dyre, and S.C. Glotzer, Phys. A. **270**, 301 (1999).

[49] P.G. Debenedetti, F.H. Stillinger, and M.S. Shell, J. Phys. Chem B **107**, 14434 (2003).

[50] Y. Pomeau and P. Resibois, Phys. Rep. **19**, 63 (1975).